\documentclass[conference]{IEEEtran}
\usepackage{amsmath,amsfonts}
\usepackage{subfigure}
\usepackage{float}

\usepackage{graphicx}

\begin{document}

\title{An Improved LR-aided K-Best Algorithm for MIMO Detection}
\author{\IEEEauthorblockN{Qi~Zhou and Xiaoli~Ma}
\IEEEauthorblockA{School of Electrical and Computer Engineering\\
Georgia Institute of Technology\\
Atlanta, Georgia 30332\\
Emails: \{qzhou32, xiaoli\}@ece.gatech.edu}}


\maketitle

\begin{abstract}
Recently, lattice reduction (LR) technique has caught great attention for multi-input multi-output (MIMO) receiver because of its low complexity and high performance. However, when the number of antennas is large, LR-aided linear detectors and successive interference cancellation (SIC) detectors still exhibit considerable performance gap to the optimal maximum likelihood detector (MLD). To enhance the performance of the LR-aided detectors, the LR-aided K-best algorithm was developed at the cost of the extra complexity on the order $\mathcal{O}(N_t^2 K + N_t K^2)$, where $N_t$ is the number of transmit antennas and $K$ is the number of candidates. In this paper, we develop an LR-aided K-best algorithm with lower complexity by exploiting a priority queue. With the aid of the priority queue, our analysis shows that the complexity of the LR-aided K-best algorithm can be further reduced to $\mathcal{O}(N_t^2 K + N_t K {\rm log}_2(K))$.  The low complexity of the proposed LR-aided K-best algorithm allows us to perform the algorithm for large MIMO systems (e.g., 50x50 MIMO systems) with large candidate sizes. Simulations show that as the number of antennas increases, the error performance approaches that of AWGN channel.
\end{abstract}

\noindent

\section{Introduction}
Designing reliable and computationally efficient multi-input multi-output (MIMO) detectors has been a longstanding challenge for wireless MIMO communications. Despite its optimal error performance, the maximum likelihood detector requires considerably high complexity, especially when the number of antennas is large \cite{Jalden2005}. In contrast, linear detectors (LDs) and successive interference cancellation (SIC) detectors require polynomial complexity but suffer from significantly degraded error performance. Recently, to improve the error performance of LDs and SIC detectors, lattice reduction (LR)-aided detection is proposed \cite{windpassinger2003low, Ma2008,Gan2009}, which show that LR-aided LDs can achieve the same diversity as the MLD.

Although significant performance improvement for LR-aided LDs and SIC detectors is found, the LR-aided detectors still exhibit considerable performance loss to the MLD. In addition, as the number of antennas increases, the gap between the LR-aided detectors and the MLD increases significantly \cite{ling2011proximity}. To further bridge the gap, the LR-aided K-best algorithm is proposed in \cite{qi2007lattice, shabany2008application}. Compared to the conventional K-best, the LR-aided K-best algorithm has no boundary information about the symbols in the lattice-reduced domain, i.e., the possible children for each layer can be infinite. To find the $K$ best partial candidates from the infinite children set, the algorithm in \cite{qi2007lattice} replaces the infinite set with a finite subset of the children. To reduce the complexity of generating the subset, study in \cite{shabany2008application} develops an on-demand child expansion based on the Schnorr-Euchner (SE) strategy.

In this paper, we further reduce the complexity of finding the $K$ best partial candidates from the infinite children set for each layer by exploiting the on-demand child expansion and a priority queue. We find that, with the aid of the priority queue, the complexity can be reduced from $\mathcal{O}(N_t^2 K + N_t K^2)$ in \cite{qi2007lattice, shabany2008application} to $\mathcal{O}(N_t^2 K + N_t K \log_2(K))$. The lower complexity in terms of the candidate sizes $K$ allows us to perform MIMO with large candidate sizes, resulting higher error performance. Our simulations show that, combined with the minimum-mean-square-error (MMSE) regularization, the proposed LR-aided K-best algorithm can achieve near-optimal error performance for large MIMO systems with large constellation sizes (e.g., 50x50 MIMO with 256QAM).

The rest of the paper is organized as follows. Section \ref{sec:signalmodel} introduces the system model and LR-aided detectors. Section \ref{sec:lrkbest} discusses the proposed LR-aided K-best algorithm. Section \ref{sec:numres} shows the numerical results. Section \ref{sec:concl} concludes the paper.

\textit{Notation:} Superscript $^T$ denotes the transpose. The real and imaginary parts of a complex number are denoted as ${\mathfrak{R}}[\cdot]$ and ${\mathfrak{I}}[\cdot]$. Upper- and lower-case boldface letters indicate matrices and column vectors, respectively. ${{A}}_{i,k}$ indicates the $(i,k)$th entry of matrix ${\mathbf{A}}$. ${\mathbf{I}}_N$ denotes the $N \times N$ identity matrix, $\mathbf{0}_{N \times L}$ is the $N \times L$ matrix with all entries zero, and $\mathbf{1}_{N \times L}$ is the $N \times L$ matrix with all entries one. ${\mathbb{Z}}$ is the integer set. $E\{ \cdot \}$ denotes the statistical expectation. $\| \cdot \|$ denotes the $2$-norm.

\section{System Model}\label{sec:signalmodel}
The transmission model of a V-BLAST MIMO system with $N_t$ transmit antennas and $N_r$ receive antennas is
\begin{equation}\label{eq:model}
{\mathbf{y}}^c = {\mathbf{H}}^c {\mathbf{s}}^c + \mathbf{w}^c,
\end{equation}
where $\mathbf{s}^c = [s^c_{1}, s^c_{2}, \cdots, s^c_{N_t}]^T, (s^c_i \in {\mathcal{S}}^c)$ is the complex information symbol vector with ${\mathcal{S}}^c$ being a constellation set of QAM, ${\mathbf{H}}^c$ is an $N_r \times N_t, (N_r \geq N_t)$ complex channel matrix, whose entries are modeled as independent and identically distributed (i.i.d.)  complex Gaussian variables with zero means and unit variances, ${\mathbf{y}}^c = [y^c_1, y^c_2, \cdots, y^c_{N_r}]^T$ is the received signal vector, and ${\mathbf{w}^c} = [w^c_{1}, w^c_{2}, \cdots, w^c_{N_r}]^T$ is the complex additive white Gaussian noise (AWGN) vector with zero mean and covariance $N_0 {\mathbf{I}}_{N_r}$.

Given the complex signal model in \eqref{eq:model}, the equivalent real signal model is
\begin{align}
\begin{bmatrix}
{\mathfrak{R}}[{\mathbf{y}}^c] \\
{\mathfrak{I}}[{\mathbf{y}}^c]
\end{bmatrix} & = \begin{bmatrix} {\mathfrak{R}}[\mathbf{H}^c] & -{\mathfrak{I}}[\mathbf{H}^c] \\
{\mathfrak{I}}[\mathbf{H}^c] & {\mathfrak{R}}[\mathbf{H}^c] \end{bmatrix} \begin{bmatrix}
{\mathfrak{R}}[{\mathbf{s}}^c] \\
{\mathfrak{I}}[{\mathbf{s}}^c]
\end{bmatrix} + \begin{bmatrix}
{\mathfrak{R}}[{\mathbf{w}}^c] \\
{\mathfrak{I}}[{\mathbf{w}}^c]
\end{bmatrix} \nonumber
\\
\mathbf{y} & = \mathbf{H} \mathbf{s} + \mathbf{w}, \label{eq:sigmodr}
\end{align}
where $\mathbf{s} = [s_1, s_2, \cdots, s_{2 N_t}]^T$ with $s_i \in \mathcal{S}$ and $\mathcal{S}$ is the constellation set of PAM as $\{ -\sqrt{M} + 1, -\sqrt{M} + 3, \cdots, \sqrt{M} - 1\}$.

\subsection{LR-aided Detectors}
Given the model in Eq. \eqref{eq:sigmodr}, the MLD is
\begin{equation}\label{eq:mld}
{\mathbf{\hat{s}}}^{\mathrm{ML}} = \arg \min_{\mathbf{\tilde{s}} \in {\mathcal{S}}^{2 N_t}} \| {\mathbf{y}} - {\mathbf{H}} {\mathbf{\tilde{s}}} \|^2,
\end{equation}
which is generally non-deterministic polynomial hard (NP-hard). To reduce the high complexity of the MLD, the LR-aided detection is proposed.

Since the LR-aided detection only works for infinite lattice, we first remove the boundary constraints in Eq. \eqref{eq:mld} and obtain the following relaxed problem
\begin{equation}\label{eq:nld}
{\mathbf{\hat{s}}} = \arg \min_{ \mathbf{\tilde{s}} \in {\mathcal{U}}^{2 N_t}} \| {\mathbf{y}} - {\mathbf{H}} {\mathbf{\tilde{s}}} \|^2,
\end{equation}
where $\mathcal{U}$ is the unconstrained constellation set as $\{ \cdots, -3, -1, 1, 3, \cdots \}$. Since ${\mathbf{\hat{s}}}$ may not be a valid QAM symbol, a quantization step can be applied
\begin{equation}
{\mathbf{\hat{s}}}^{\mathrm{NLD}} = \mathcal{Q}( {\mathbf{\hat{s}}} ),
\end{equation}
where ${\mathcal{Q}}(\cdot)$ the symbol-wise quantizer to the constellation set $\mathcal{S}$.

In essence, the unconstrained detection in Eq. \eqref{eq:mld} is the naive lattice detection (NLD) studied in \cite{jalden2010dmt, taherzadeh2010limitations}. The closest point search algorithm \cite{Agrell2002} (e.g., sphere decoding algorithm) can find the optimal solution to \eqref{eq:nld}. However, one issue of the NLD is that it is not diversity-multiplexing tradeoff (DMT) optimal in general \cite{taherzadeh2010limitations}, i.e., the NLD is suboptimal in terms of diversity. To achieve the DMT optimality, the regularized lattice decoding is proposed as \cite{jalden2010dmt}
\begin{align}
{\mathbf{\hat{s}}} & = \arg \min_{ \mathbf{\tilde{s}} \in {\mathcal{U}}^{2 N_t}} \| {\mathbf{y}} - {\mathbf{H}} {\mathbf{\tilde{s}}} \|^2 + \frac{N_0}{2 \sigma^2_s} \| \mathbf{\tilde{s}} \|^2 \nonumber \\
& = \arg \min_{ \mathbf{\tilde{s}} \in {\mathcal{U}}^{2 N_t}} \| {\mathbf{\bar{y}}} - {\mathbf{\bar{H}}} {\mathbf{\tilde{s}}} \|^2 \label{eq:nldmmse}
\end{align}
where we adopt the MMSE regularization, ${E}\{ {\mathbf{s}} {\mathbf{s}}^{T} \} = \sigma^2_{s} {\mathbf{I}}$, and $\mathbf{\bar{H}}$ and $\mathbf{\bar{y}}$ are the MMSE-extended matrix and the extended received signal vector as
\begin{equation}\label{eq:nldmmseex}
\mathbf{\bar{H}} = \begin{bmatrix} \mathbf{H} \\ \sqrt{\frac{N_0}{2 \sigma^2_s} } \mathbf{I}_{2 N_t}\end{bmatrix}, \qquad \mathbf{\bar{y}} = \begin{bmatrix} \mathbf{{y}} \\ \mathbf{0}_{2 N_t \times 1} \end{bmatrix}.
\end{equation}
We call the detection in \eqref{eq:nldmmse} the ``NLD with MMSE.'' Unless stated otherwise, in the following, we focus on solving the NLD with MMSE in Eq. \eqref{eq:nldmmse}, but the discussed LR-aided K-best algorithm is also applicable to the NLD in Eq. \eqref{eq:nld}.

To solve the NLD with MMSE in Eq. \eqref{eq:nldmmse} with lower complexity, the LR-aided detection performs LR on the matrix $\mathbf{\bar{H}}$ to obtain a more ``orthogonal'' matrix $\mathbf{\tilde{H}} = \mathbf{\bar{H}} \mathbf{T}$, where $\mathbf{T}$ is a unimodular matrix, such that all the entries of ${\mathbf{T}}$ are integers, and the determinant of ${\mathbf{T}}$ is $\pm 1$. Given $\mathbf{\tilde{H}}$ and $\mathbf{T}$, the NLD with MMSE becomes
\begin{equation}\label{eq:nldmmselr}
{\mathbf{\hat{s}}}  = 2 \mathbf{T} \arg \min_{ \mathbf{\tilde{z}} \in {\mathbb{Z}}^{2 N_t}} \| {\mathbf{\tilde{y}}} - {\mathbf{\tilde{H}}} {\mathbf{\tilde{z}}} \|^2 + \mathbf{1}_{2 N_t \times 1},
\end{equation}
where $\mathbf{\tilde{y}}$ is the received signal vector after shifting and scaling as $(\mathbf{\bar{y}} - \mathbf{\bar{H}} \mathbf{1}_{2 N_t \times 1}) / 2$ and $\mathbf{\tilde{s}} = 2 \mathbf{T} \mathbf{\tilde{z}} + \mathbf{1}_{2 N_t \times 1}$. Since $\mathbf{\tilde{H}}$ is more ``orthogonal,'' the closest point search algorithm based on $\mathbf{\tilde{H}}$ can enjoy much lower complexity compared to that based on $\mathbf{\bar{H}}$ in Eq. \eqref{eq:nldmmseex} \cite{Agrell2002}. However, since the problem in Eq. \eqref{eq:nldmmselr} is NP-hard, the complexity of the closest point search is still considerably high when $N_t$ is large. To achieve low-complexity detection, the LR-aided MMSE-SIC detector finds a sub-optimal solution to \eqref{eq:nldmmselr} with degraded error performance.

\section{LR-aided K-Best Algorithm}\label{sec:lrkbest}
To enhance the performance of the LR-aided MMSE-SIC detector, the LR-aided K-best algorithm \cite{qi2007lattice, shabany2008application} is proposed to find a ``better'' sub-optimal solution to \eqref{eq:nldmmselr}.

\begin{table}[b]
\centering
\begin{tabular}{l  l}
\hline
\multicolumn{2}{l}{\textbf{Input:} Channel matrix $\mathbf{H}$, received signal vector $\mathbf{y}$, candidate size $K$ } \\
\multicolumn{2}{l}{
\textbf{Output:} Hard-output estimate $\mathbf{\hat{s}}$}  \\
\hline
\textbf{(1)} & Obtain $\mathbf{\bar{H}}$ and $\mathbf{\bar{y}}$ in Eq. \eqref{eq:nldmmseex}
 \\
\textbf{(2)} & $[\mathbf{\tilde{H}}, \mathbf{T}] = {\rm LR}(\mathbf{\bar{H}})$ \\
\textbf{(3)} & $[\mathbf{Q}, \mathbf{R}] = {\rm QR}(\mathbf{\tilde{H}})$ \\
\textbf{(4)} & $\mathbf{\breve{y}} = {\mathbf{Q}}^T (\mathbf{\bar{y}} - \mathbf{\bar{H}} \mathbf{1}_{2 N_t \times 1}) / 2$ \\
\textbf{(5)} & $\mathbf{z}^{(2 N_t + 1)}_1 = [], cost^{(2 N_t + 1)}_1 = 0$, $len = 1$ \\
\textbf{(6)} & For $n = 2 N_t$ downto $1$ \\
\textbf{(7)} & \qquad $[ \{ \mathbf{z}^{(n)}_k \}_{k = 1}^{K}, \{ cost^{(n)}_k \}_{k = 1}^{K}]  =$ \\
& \qquad \qquad ${\rm Find\_Kbest\_Children}( \{ \mathbf{z}^{(n + 1)}_k \}_{k = 1}^{len}, \{ cost^{(n + 1)}_k \}_{k = 1}^{len})$\\
\textbf{(8)} & \qquad $len = K$ \\
\textbf{(9)} & End for \\
 \textbf{(10)}& Output $\mathbf{\hat{s}} = \arg \min_{\mathbf{\tilde{s}}_k = \mathcal{Q}(2 \mathbf{T} \mathbf{z}_k^{(1)} + \mathbf{1}_{2 N_t \times 1})} \| \mathbf{y} - \mathbf{H} \mathbf{\tilde{s}}_k\|^2$ \\
\hline
\end{tabular}
\caption{A general description of the LR-aided K-best algorithm.}
\label{alg:lrkbest}
\end{table}

\begin{table*}[t]
\centering
\begin{tabular}{l l l}
\hline
\multicolumn{3}{l}{\textbf{Input:} $len$ partial candidates of the $(n + 1)$st layer $\{ \mathbf{z}^{(n + 1)}_k \}_{k = 1}^{len}$ with their costs $\{ cost^{(n + 1)}_k \}_{k = 1}^{len}$ } \\
\multicolumn{3}{l}{
\textbf{Output:} $K$ partial candidates of the $n$th layer $\{ \mathbf{z}^{(n)}_k \}_{k = 1}^{K}$ with their costs $\{ cost^{(n)}_k \}_{k = 1}^{K}$ } \\
\hline
Line no. & Description & Complexity \\
\hline
\textbf{(1)} &  For $i = 1$ to $len$\\
\textbf{(2)} &  \qquad $r_i = \breve{y}_n - \sum_{j = n + 1}^{2 N_t} R_{n, j} z^{(n + 1)}_{i,j}$ & $\mathcal{O}(N_t)$  \\
\textbf{(3)} &  \qquad $z_i = \lceil r_i / R_{n,n} \rfloor$ & $\mathcal{O}(1)$\\
\textbf{(4)} &  \qquad $child_i = [z_i , (\mathbf{z}^{(n + 1)}_i)^T]^T$ &  $\mathcal{O}(1)$ or $\mathcal{O}(N_t)$ \\
\textbf{(5)} &  \qquad $childcost_i = cost^{(n + 1)}_i + (r_i - R_{n,n} z_i)^2 $ & $\mathcal{O}(1)$ \\
\textbf{(6)} &  \qquad $step_i = {\rm sgn}(r_i / R_{n,n} - z_i )$ & $\mathcal{O}(1)$\\
\textbf{(7)} &  End for \\
\textbf{(8)} &  Initialize a priority queue $q$ with $\{childcost_i \}_{i = 1}^{len}$ as the keys & $\mathcal{O}(K)$\\
\textbf{(9)} &  For $k = 1$ to $K$ \\
\textbf{(10)} &  \qquad Find the index $i$ associated with the minimum key in $q$ & $\mathcal{O}(1)$\\
\textbf{(11)} &  \qquad $\mathbf{z}^{(n)}_k = child_i$ & $\mathcal{O}(1)$  or $\mathcal{O}(N_t)$ \\
\textbf{(12)} &  \qquad $cost^{(n)}_k = childcost_i$ & $\mathcal{O}(1)$ \\
\textbf{(13)} &  \qquad $z_i = z_i + step_i$ & $\mathcal{O}(1)$ \\
\textbf{(14)} &  \qquad $child_i = [z_i , (\mathbf{z}^{(n + 1)}_i)^T]^T$ & $\mathcal{O}(1)$\\
\textbf{(15)} &  \qquad $childcost_i = cost^{(n + 1)}_i + (r_i - R_{n,n} z_i)^2 $ & $\mathcal{O}(1)$\\
\textbf{(16)} &  \qquad $step_i = -step_i -{\rm sgn}(step_i)$ & $\mathcal{O}(1)$\\
\textbf{(17)} &  \qquad Update $q$ using $childcost_i$ as the new key & $\mathcal{O}(\log_2(K))$ \\
\textbf{(18)} &  End for \\
\textbf{(19)} &  Output $\{ \mathbf{z}^{(n)}_k \}_{k = 1}^{K}, \{ cost^{(n)}_k \}_{k = 1}^{K}$ \\
\hline
\end{tabular}
\caption{The proposed ${\rm Find\_Kbest\_Children}()$ subroutine for the LR-aided K-best algorithm.}
\label{alg:findkbest}
\end{table*}

The LR-aided K-best algorithm first performs QR decomposition on $\mathbf{\tilde{H}} = \mathbf{Q} \mathbf{R}$, where $\mathbf{Q}$ is a $2 (N_r + N_t) \times 2 N_t$ orthonormal matrix and $\mathbf{R}$ is a $2 N_t \times 2 N_t$ upper triangular matrix. Then, the problem in \eqref{eq:nldmmselr} is reformulated as
\begin{equation}\label{eq:nldmmselrQR}
{\mathbf{\hat{s}}}  = 2 \mathbf{T} \arg \min_{ \mathbf{\tilde{z}} \in {\mathbb{Z}}^{2 N_t}} \| \mathbf{\breve{y}} - {\mathbf{R}} {\mathbf{\tilde{z}}} \|^2 + \mathbf{1}_{2 N_t \times 1}.
\end{equation}
where $\mathbf{\breve{y}} = \mathbf{Q}^T {\mathbf{\tilde{y}}}$.

Next, the LR-aided K-best algorithm performs the breadth-first search from the $2 N_t$th layer to the $1$st layer. For each layer (e.g., the $n$th layer), the algorithm computes the $K$ best partial candidates $[\mathbf{z}^{(n)}_1, \mathbf{z}^{(n)}_2, \cdots, \mathbf{z}^{(n)}_K]$, i.e., the $K$ partial candidates with the minimum costs among all the children of the $K$ partial candidates $[\mathbf{z}^{(n + 1)}_1, \mathbf{z}^{(n + 1)}_2, \cdots, \mathbf{z}^{(n + 1)}_K]$ in the previous $(n + 1)$st layer, where a partial candidate $\mathbf{z}^{(n)}_i$ in the $n$th layer is $[{z}^{(n)}_{i, n}, \cdots, {z}^{(n)}_{i, 2 N_t}]^T$ and the cost associated with the partial candidate is
\begin{equation}
cost^{(n)}_i = \sum_{j = n}^{2 N_t} (\breve{y}_j - \sum_{k = j}^{2 N_t} R_{j,k} z^{(n)}_{i, k})^2.
\end{equation}
We call a partial candidate of the $n$th layer $\mathbf{z}^{(n)}_i$ a child of a partial candidate of the $(n + 1)$st layer $\mathbf{z}^{(n + 1)}_j$ if and only if $\mathbf{z}^{(n)}_i = [z^{(n)}_{i, n}, (\mathbf{z}^{(n + 1)}_j)^T]^T, z^{(n)}_{i,n} \in \mathbb{Z}$ holds.

A general description of the LR-aided K-best algorithm is given in Table \ref{alg:lrkbest}. Note that only one partial candidate is in the $(2 N_t + 1)$st layer, where $\mathbf{z}^{(2 N_t + 1)}_1$ represents the root node.

From Table \ref{alg:lrkbest}, the key task of the LR-aided K-best algorithm is how to efficiently find the $K$ best partial candidates of each $n$th layer from all the children of the partial candidates of the previous $(n + 1)$st layer. However, different from the K-best algorithm in s-domain, in which the number of children is finite due to the bounded constellation set $\mathcal{S}$, each partial candidate in the LR-aided K-best algorithm has infinite possible children because no information about the boundary of $\mathbf{z}$ is available.

To address the infinite children issue, the algorithm in \cite{qi2007lattice} approximates the infinite children set with a finite set with $N K$ children by expanding only $N$ best children for each partial candidate of the $(n + 1)$st layer. The algorithm then chooses the top $K$ partial candidates for the $n$th layer from the $N K$ children. Note that, when $N = K$, the $K^2$ children themselves contain at least $K$ best partial candidates among all the children of the $(n + 1)$st layer, and thus, the algorithm in \cite{qi2007lattice} becomes an exact one. To further reduce the number of node expansions, the algorithm in \cite{shabany2008application} employs the Schnorr-Euchner (SE) strategy to perform an on-demand child expansion, where a child is expanded if only if all its better siblings are expanded and chosen as the partial candidates of the $n$th layer. Although significant reduction on the node expansions is achieved in \cite{shabany2008application}, both algorithms in \cite{qi2007lattice, shabany2008application} require at least $\mathcal{O}(N_t K + K^2)$ operations to find the exact $K$ best partial candidates for each layer.

In this paper, we further reduce the complexity of finding the exact $K$ best partial candidates for each layer to $\mathcal{O}(N_t K + K \log_2(K))$ by exploiting the on-demand child expansion and a priority queue. The pseudo code of the method is given in Table \ref{alg:findkbest}. Compared to the algorithm in \cite{shabany2008application}, the proposed algorithm employs a priority queue instead of a brute-force method in \cite{shabany2008application} to find a child with the minimum cost in line 10. The priority queue can be implemented by a heap, which requires $\mathcal{O}(1)$ operations to find the child with the minimum cost, $\mathcal{O}(\log_2(K))$ operations to maintain the heap if a key is changed (line 17), and $\mathcal{O}(K)$ operations to initialize the heap with $K$ elements (line 8). Thus, the overall complexity of the proposed method is $\mathcal{O}(N_t K + K \log_2(K))$, which is considerably lower than $\mathcal{O}(N_t K + K^2)$ in \cite{qi2007lattice, shabany2008application} when $K$ is large. Note that, the complexity of lines 4 and 11 generally relies on the data structure of the implementation and is at most on the order $\mathcal{O}(N_t)$ by using a memory copy method.

\section{Numerical Results}\label{sec:numres}
This section demonstrates the performance of the proposed LR-aided K-best algorithm for MIMO systems. For the LR algorithm, we adopt the LLL algorithm \cite{lenstra1982} with reduction quality parameter $\delta = 0.99$. The SNR is defined as the received information bit energy versus noise variance.

\begin{figure}[t]
\centering
\includegraphics[width=0.90\linewidth]{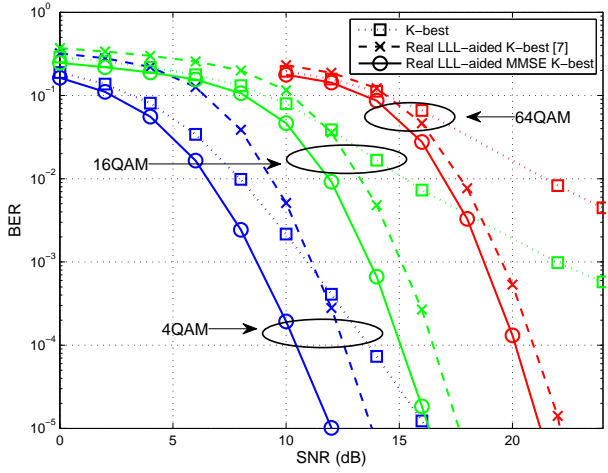}
\caption{Performance comparisons of different K-best algorithms for 10x10 MIMO systems with $K = 5$ and different QAMs.}\label{fig:diff_kbests}
\end{figure}

Fig. \ref{fig:diff_kbests} displays the performance comparisons among the three K-best algorithms: {\em i}) the K-best algorithm in s-domain; {\em ii}) the LLL-aided K-best algorithm for the NLD proposed in \cite{shabany2008application}; and {\em iii}) the LLL-aided K-best algorithm for the NLD with MMSE in Eq. \eqref{eq:nldmmselr}, which is called the ``LLL-aided MMSE K-best algorithm.'' From the figure, it is clear to see that, given a fixed number of $K$, both the LLL-aided K-best algorithm and the LLL-aided MMSE K-best algorithm achieve significant gain over the K-best algorithm at high SNR, especially for high order QAMs. In addition, the LLL-aided MMSE K-best algorithm shows better performance than the LLL-aided K-best algorithm, especially for 4QAM, where the LLL-aided MMSE K-best algorithm has about $2.5$ dB gain over the LLL-aided K-best algorithm at BER=$10^{-5}$.

\begin{figure}[t]
\centering
\includegraphics[width=0.90\linewidth]{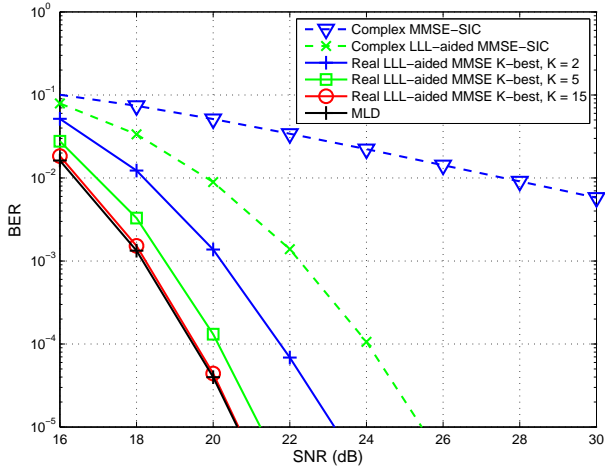}
\caption{Performance comparisons of the LLL-aided MMSE K-best algorithm for a 10x10 MIMO system with 64QAM and different $K$.}\label{fig:kbest_N10_64QAM}
\end{figure}

Fig. \ref{fig:kbest_N10_64QAM} depicts the performance comparisons among the LLL-aided MMSE K-best algorithm, the MLD, the complex MMSE-SIC, and the complex LLL-aided MMSE-SIC. We have the following observations: {\em i}) The LLL-aided MMSE-SIC obtains significant performance improvement over the MMSE-SIC, whose diversity is $1$. {\em ii}) The LLL-aided MMSE K-best algorithm further enhances the performance of the LLL-aided MMSE-SIC, where the LLL-aided MMSE K-best algorithm with $K = 2$ has more than $2$ dB gain over the LLL-aided MMSE-SIC. {\em iii}) By increasing the number of candidates $K$, the performance of the LLL-aided MMSE K-best algorithm approaches that of the MLD, and when $K = 15$, the LLL-aided MMSE K-best can achieve almost the same performance as the MLD.

\begin{figure}[t]
\centering
\includegraphics[width=0.90\linewidth]{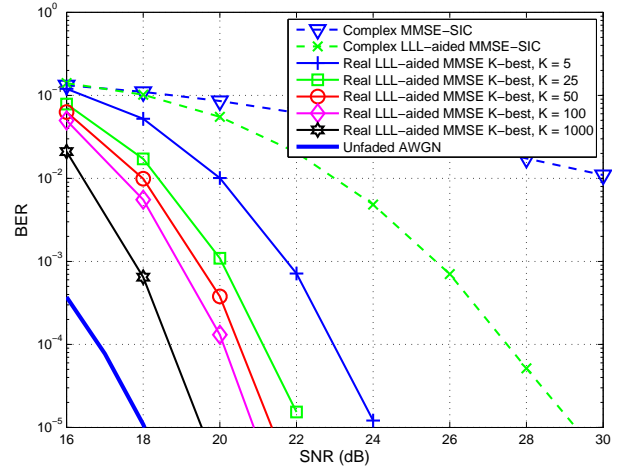}
\caption{Performance comparisons of the LLL-aided MMSE K-best algorithm for a 32x32 MIMO system with 64QAM and different $K$.}\label{fig:kbest_N32_64QAM}
\end{figure}

Fig. \ref{fig:kbest_N32_64QAM} illustrates the performance of the different detectors for a 32x32 MIMO system with 64QAM. First, we observe that, compared to that in Fig. \ref{fig:kbest_N10_64QAM}, the performance of the LLL-aided MMSE-SIC degrades significantly (e.g., the LLL-aided MMSE-SIC requires around $25.5$ dB for 10x10 MIMO and around $29.3$ dB for 32x32 MIMO at BER = $10^{-5}$). Second, the LLL-aided MMSE K-best algorithm significantly improves the performance of the LLL-aided MMSE-SIC, which has about $5$ db loss to the LLL-aided MMSE K-best algorithm with $K = 5$. Furthermore, the LLL-aided MMSE K-best algorithm with $K = 1000$ exhibits less than $2$ dB loss to the unfaded AWGN bound at BER = $10 ^ {-5}$.

\begin{figure}[t]
\centering
\includegraphics[width=0.90\linewidth]{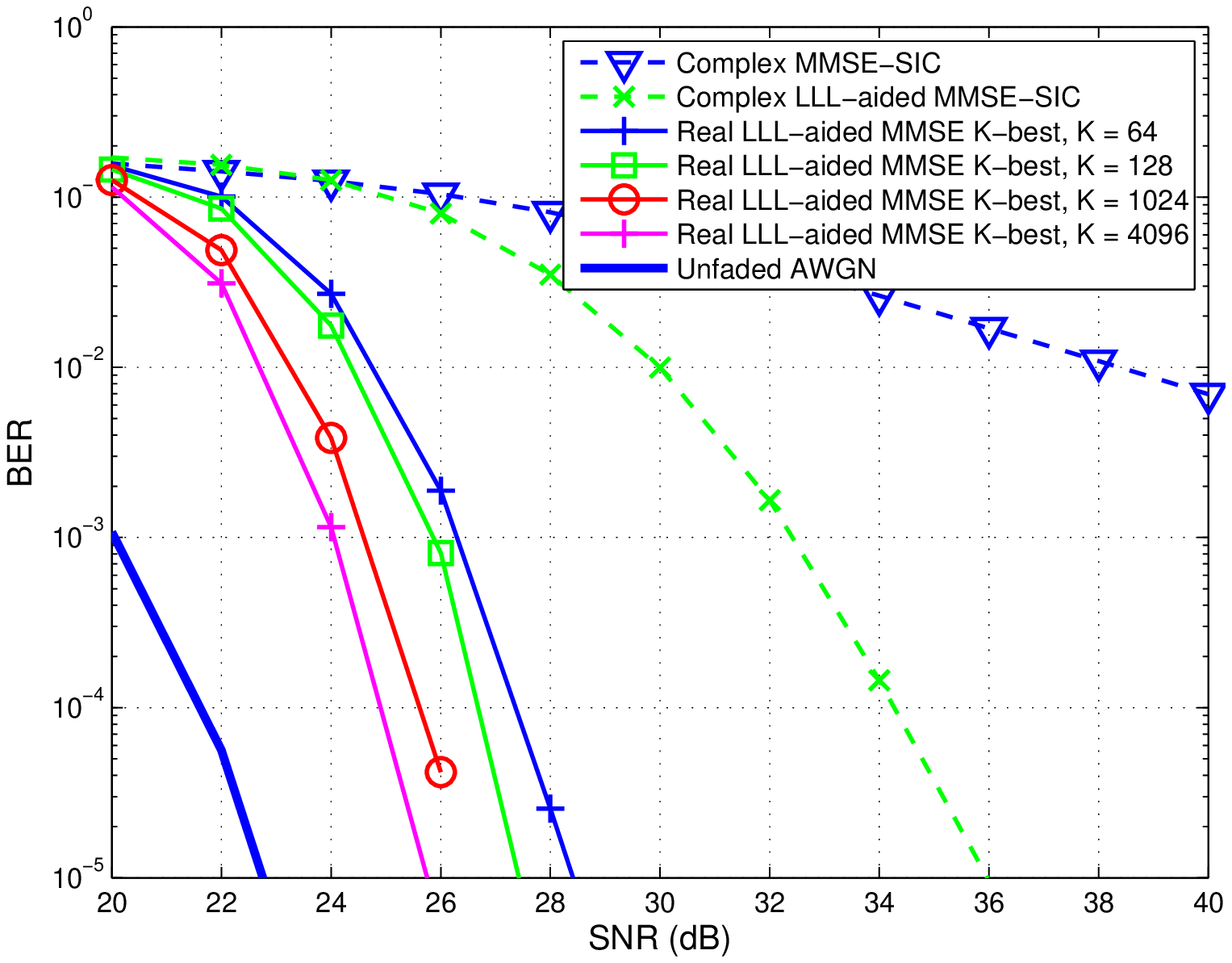}
\caption{Performance comparisons of the LLL-aided MMSE K-best algorithm for a 50x50 MIMO system with $256$QAM and different $K$.}\label{fig:kbest_N50_256QAM}
\end{figure}

To further investigate the performance of the LLL-aided MMSE K-best algorithm for large MIMO systems with large constellation sizes, we compare the performance of the different detectors for a 50x50 MIMO system with 256QAM in Fig. \ref{fig:kbest_N50_256QAM}. From the figure, we find that the proposed LR-aided MMSE K-best algorithm with $K = 128$ and $K = 4096$ can achieve about $5$ dB and $3$ dB losses to the unfaded AWGN bound at BER = $10^{-5}$, respectively.

\section{Concluding Remarks}\label{sec:concl}
In this paper, we proposed an LR-aided K-best algorithm by exploiting an on-demand child expansion strategy and a priority queue. Our complexity analysis showed that the complexity of the proposed algorithm is $\mathcal{O}(N_t^2 K + N_t K \log_2(K))$, which is lower than that of the exiting ones, especially for large candidate size $K$. Our simulations demonstrated that, with large number of $K$, the proposed LR-aided K-best algorithm can achieve near-optimal performance for large MIMO systems.

\section*{Acknowledgment}
Part of the work is supported by ARO Grant DAAD W911NF-11-1-0542.

{
\bibliography{IEEEabrv,qi}
\bibliographystyle{IEEEtran}
}

\end{document}